\documentclass[pra, twocolumn, showpacs, eqsecnum]{revtex4}
\usepackage{amsmath}
\usepackage{amssymb}
\usepackage{graphicx}
\usepackage{stackrel}

\newcommand{\coeffM}{m}
\newcommand{\thouless}{\xi_L}

\newcommand{\order}{\mathcal{O}}
\newcommand{\rmi}{\mathrm{i}}
\newcommand{\rms}{\mathrm{s}}
\renewcommand{\rmp}{\mathrm{p}}
\newcommand{\rmd}{\mathrm{d}}

\newcommand{\FCL}{F_\text{CL}}
\newcommand{\fCL}{f_\text{CL}}
\newcommand{\fD}{f_{D}}
\newcommand{\DispersionD}{\mathcal{D}}
\newcommand{\rCL}{\rho_\text{CL}}
\newcommand{\rD}{\rho_{D}}
\newcommand{\intMD}{\mathcal{M}}

\newcommand{\be}{\begin{equation}}
\newcommand{\ee}{\end{equation}}

\begin{document}

\title{Casimir-Foucault interaction: Free energy and entropy at low temperature}

\author{Francesco Intravaia}
\affiliation{Theoretical Division, MS B213, Los Alamos National Laboratory, Los Alamos, NM 87545, USA}
\author{Simen {\AA}. \surname{Ellingsen}}
\affiliation{Department of Energy and Process Engineering, Norwegian
University of Science and Technology, N-7491 Trondheim, Norway}
\author{Carsten Henkel}
\affiliation{Institut f\"{u}r Physik und Astronomie, Universit\"{a}t Potsdam, Karl-Liebknecht-Str.\ 24/25, 14476 Potsdam, Germany}

\begin{abstract}
It was recently found that thermodynamic anomalies which arise in the Casimir effect between metals described by the Drude model can be attributed to the interaction of fluctuating Foucault (or eddy) currents [Phys.\ Rev.\  Lett.\ {\bf 103}, 130405 (2009)]. We show explicitly that the two leading terms of the 
low-temperature correction to the Casimir free energy of interaction between
two plates, are identical to those pertaining to the Foucault current interaction alone, up to a correction which is very small for good metals. 
Moreover, a mode density along real frequencies is introduced, showing that 
the Casimir free energy, as given by the Lifshitz theory, separates in a natural manner
in contributions from eddy currents and propagating cavity modes, respectively.
The latter have long been known to be of little importance to the low-temperature Casimir anomalies. 
This convincingly demonstrates that  eddy
current modes are responsible for the large temperature correction to the Casimir effect between Drude metals, predicted by the Lifshitz theory, but not observed 
in experiments.
\end{abstract}

\date{03 May 2010}

\maketitle

\section{Introduction}

For a decade the finite-temperature correction to the Casimir force\cite{casimir48} between parallel metal plates has been a topic of intense investigation and debate. Describing the metals by a standard Drude model 
\be
  \varepsilon(\omega) = 1 - \frac{\Omega^2 }{
  \omega(\omega + {\rm i}\gamma)},
  	\label{eq:drude}
\ee
where $\Omega$ is the plasma frequency and where the relaxation frequency $\gamma$ does not vanish at $T=0$, 
the Lifshitz theory implies that the temperature dependence
is considerably different from perfect reflectors\cite{bostrom00}: a significant
thermal contribution is predicted already at distances shorter than the Wien
wavelength $\hbar c / (2\pi k_B T)$, on the one hand,
and there is a difference of a factor $\frac12$ in the large-distance limit,
on the other.
Puzzlingly, such a large temperature dependence is not found in recent precision experiments at Purdue \cite{decca07}.
For reviews of the thermal debate around the Casimir effect, cf.\ \cite{brevik06, BookBordag09} and references therein.

The thermodynamics of the Casimir effect has been of particular interest 
in this context. For metals described by (\ref{eq:drude}), the Gibbs-Helmholtz 
free energy of the Casimir interaction is non-monotonous as a function of
temperature, leading to a negative Casimir entropy in a large temperature 
range \cite{brevik06}. Moreover, if $\gamma$ vanishes faster than linear as
the temperature $T \to 0$, the Casimir entropy remains nonzero in this limit;
this was argued to violate Nernst's theorem, the third law of 
thermodynamics \cite{klimchitskaya01}.

Recently two of the present authors investigated the contribution to the Casimir force from Johnson-Nyquist noise, focusing on specific solutions of the Maxwell 
equations for the two-plate set-up, namely purely dissipative (i.e., overdamped) 
modes which are physically Foucault current or `eddy current' modes \cite{intravaia09}. [A related 
investigation with a simplified
model is due to Bimonte \cite{bimonte07}.] It was shown that the eddy current contribution alone accounts for the apparently anomalous thermodynamics of 
the Casimir effect. The non-vanishing entropy that
appears when
first $\gamma\to 0$ and then $T \to 0$ (taken in this order), is due to an infinite degeneracy of quasi-static Foucault current states, a glass-like situation for which the Nernst theorem does not apply \cite{intravaia09b}.
The situation is closely analogous to that of a free particle coupled to a heat bath \cite{ingold09}, which is essentially in its high temperature limit for any nonzero temperature when no damping is present, and for which the Nernst theorem is satisfied for a fixed friction rate \cite{hanggi06}. The apparent thermodynamical anomaly in the Casimir context was investigated in detail
in Refs.\ \cite{intravaia08,ellingsen08b,ellingsen09}.
It is now established that the Lifshitz theory gives a low-temperature expansion 
of the 
Casimir free energy between two Drude metals in the 
form \cite{ingold09, brevik04,hoye07,ellingsen08,Brevik08b,ellingsen09}
\be\label{FCL}
  \Delta \FCL(T) = \fCL^{(2)} T^2 + \fCL^{(5/2)}T^{5/2}+...
\ee
where the free energies are split into
\be
  F(T) = F_0 + \Delta F(T),
\ee
$F_0$ being the zero temperature value. The derivation of these results
in Refs.\cite{hoye07, Brevik08b, ellingsen08}, starting from a Matsubara
sum, is quite tricky, see Sec.\ 3 of Ref.\ \cite{ellingsen09}.
They were confirmed independently using a scattering approach
by Ingold and collaborators~\cite{ingold09}.

In this paper we go one step further in showing how the 
behavior of the Casimir effect between good Drude metals is dictated 
entirely by the contribution from Foucault current modes. 
We introduce a free energy of interaction between Foucault
current modes in two Drude plates, and find it to have the same form
at low temperatures:
\be\label{eq:FD-expansion}
  \Delta F_D(T) = \fD^{(2)} T^2 + \fD^{(5/2)}T^{5/2}+...
\ee
We use throughout the subscript $D$ to denote the eddy current (or
diffusive modes) contribution to the Casimir interaction.
We are able to calculate the coefficients $\fD^{(2)}$ and $\fD^{(5/2)}$ 
and, quite remarkably, find  
\begin{subequations}\label{CoefficientRelation}
\begin{align}
  \fD^{(2)} &= \fCL^{(2)} + \order[(\gamma/\Omega)^2];\\
  \fD^{(5/2)} &= \fCL^{(5/2)}.
\end{align}
\end{subequations}
The calculations are based on an analysis of the zeros and branch cuts 
of the dispersion function for the Casimir energy, similar to previous work based
on the argument principle \cite{Davies72, Schram73}.
This analysis permits us to identify a density of states (DOS) for 
both the Foucault-current interaction and the full electromagnetic 
Casimir interaction within the Lifshitz theory.
This method reveals a close relationship between the two interactions,
and yields the results (\ref{CoefficientRelation}) in a fairly simple way,
including the correction term to $f^{(2)}$ of order 
$\gamma^2/\Omega^2$ which we calculate in the limit of good conductors.

The low-temperature expansion is valid on a temperature scale lower than
\be\label{eq:D}
  k_B T \ll \frac{ \hbar D }{ L ^2 } = \frac{\hbar \gamma c^2}{\Omega^2 L^2}
\ee
where $D$ is the diffusion coefficient of Foucault currents \cite{Jackson} and $L$ the distance between the plates. This scale 
(`Thouless energy' \cite{thouless77}) has been
identified in previous work~\cite{torgerson04, svetovoy07, ellingsen08b} and 
emerges naturally when spatially diffusive modes in two half-spaces are
coupled by electromagnetic fields across a gap of width 
$L$~\cite{intravaia09}.
It corresponds to a temperature around
20\,K for $L = 100\,{\rm nm}$ and the conductivity of gold at 
room temperature.
We shall refer frequently to this parameter in the following.

The paper is structured as follows: in section \ref{sec:FfromDOS}, we introduce a general scheme for calculating the low temperature expansion of the Gibbs-Helmholtz free energy from DOS functions, and recapitulate the DOS for the Casimir-Lifshitz and Foucault current interactions, respectively. We use a method of contour integration to derive a relation between the DOS of the two types of interaction. This provides an intuitive tool for calculating the desired expansion coefficients $f^{(2)}$ and $f^{(5/2)}$ in section \ref{sec:leadingorder}, first to leading order in the small parameter $\gamma/\Omega$, then the correction term. Various mathematical results are collected in the appendixes.

Throughout the calculation we assume the material be described by~(\ref{eq:drude}), and let 
 $\hbar = k_{\rm B} =1$. We shall use the terms eddy current and Foucault current interchangably.

\section{Mode densities}
\label{sec:FfromDOS}

\subsection{Introduction}

The Gibbs-Helmholtz free energy $F$ for a 
system with a continuous distribution of bosonic normal modes is 
related to the DOS $\rho(\omega)$ (modes per angular frequency) 
by the relation 
\begin{align}
	\Delta F( T ) 	&= 
	T \int\limits_0^{\infty} \!{\rm d}\omega\, \rho( \omega ) 
	\log(1 - {\rm e}^{-\omega / T})
\notag \\
	&=\int\limits_0^{\infty}\!{\rm d}\omega
	\frac{ \intMD( \omega ) }{ {\rm e}^{\omega / T} - 1 },
\label{eq:free-energy-integral}
\end{align}
where $ \intMD( \omega )$ is the integrated mode density:
\be\label{eq:def-M}
  \rho( \omega ) =- \partial_{\omega} \intMD( \omega ).
\ee
(We fix the integration constant with
$\intMD( 0 ) = 0.$)
The mode density $\rho( \omega )$ (per angular frequency)
specifies the physical system. (Note the difference to the density of states 
per unit energy introduced in Ref.\cite{Hanke95}.)
In the low-temperature limit,
the exponential confines the integrand to small values of $\omega$, 
and we can expand $\intMD( \omega )$ in powers of $\omega$ [see 
also Ref.\ \cite{Ford05a}]. 
Integrating termwise, each power $\omega^\nu$ of the expansion yields a contribution 
$\sim T^{\nu+1}$ according to
\begin{align}
	\int_0^\infty &\frac{\rmd \omega \, \omega^\nu}{e^{\omega/T} - 1} 
	= \Gamma(\nu+1)\zeta(\nu+1)T^{\nu+1}.
\label{eq:zeta-function-identity}
\end{align}
This method is the real-frequency analog of the method laid out 
in \cite{ellingsen09} and used in \cite{ellingsen08} where Matsubara sums
were expanded at low temperatures.
The exponential cutoff from the temperature dependence makes the procedure
considerably more straightforward here, since standard methods of asymptotic expansion are applicable.

For the two-plate geometry, $F$ is a free energy per
area and also depends on their separation $L$, with the corresponding 
pressure given by $p = - \partial F / \partial L$. 
The low-frequency expansion of the mode density ${\cal M}_D( \omega )$ 
for diffusive modes is found to be of the form
\begin{equation}
	\intMD_{D}( \omega ) \approx 
	\left[ \coeffM^{(1)}_{D}\frac{\omega}{D} +
	\coeffM^{(3/2)}_{D} L \left(\frac{\omega}{D}\right)^{3/2}\dots
	\right]
	\label{eq:small-omega-expansion}
\end{equation}
where the inverse diffusion constant $1/D$ conveniently provides the physical 
units, 
and the 
Thouless frequency $D / L^2$ gives the relevant frequency scale.
The coefficients 
$\coeffM^{(1)}_{D}$, $\coeffM^{(3/2)}_{D}$ 
are dimensionless, the first of which
relates quite obviously
to the static value of the mode density [see \eqref{eq:def-M}]:
\be
  \rho_D( 0 ) = - \frac{   \coeffM^{(1)}_{D} }{ D }
.
\ee
Applying the 
identities~(\ref{eq:zeta-function-identity}), we get the desired free energy
expansion
\begin{multline}
	\Delta F_D( T ) =
	 \zeta(2)  \frac{\coeffM^{(1)}_{D}}{D}T^2\\
	+  
	\frac{ \sqrt{\pi}\, 	\zeta(\frac{5}{2}) }{ 2 }
	\coeffM^{(3/2)}_{D} 
	\frac{ T^{5/2} L }{ D^{3/2} }
	+ \mathcal{O}( T^3 )
	\label{eq:F-overview-low-T-expansion}
\end{multline}
where $\zeta(2) = \pi^2/6$ and $\zeta(\frac{5}{2}) \approx 1.34149$.
As we calculate in Section~\ref{sec:leadingorder} below 
\begin{subequations}\label{eq:result-rhoD}
\begin{eqnarray}
	\coeffM^{(1)}_{D}  &\approx& 
	 \frac{ 2 \log 2 - 1}{ 8 \pi^2  }
	+  \frac{ \lambda (\gamma / \Omega)^2 }{ 4 \pi^2 ( L + 2 \lambda) };\label{eq:M1}
	\\
	\coeffM^{(3/2)}_{D}  &=& 
	-\frac{ \sqrt{ 2 } }{ 24\pi^2  },\label{eq:M32}
\end{eqnarray}
\end{subequations}
where an expansion for good conductors
($\gamma \ll \Omega$) has been performed,
with corrections to $\coeffM^{(1)}_{D}$ appearing at the order 
$\mathcal{O}^2(\gamma/\Omega)$. The plasma penetration depth
is defined as $\lambda = c / \Omega$.
Note that the limit $L \to \infty$ cannot be applied here, since it
conflicts with the small parameter $T L^2 / D$ in the expansion
[Eq.(\ref{eq:D})]; this
is why the scaling with $L$ in the third term on the right hand side of 
of Eq.(\ref{eq:F-overview-low-T-expansion}) is not unphysical. 
In the limit $L \to 0$, 
$\Delta F_D$ 
is nonzero and finite: this
can be attributed to the change in the bulk self-energy of the electromagnetic 
excitations of the metallic medium, as a pair of surfaces is introduced 
(the `cleavage energy' discussed by Barton \cite{Barton79}). 
The surfaces introduce boundary conditions for the fluctuating 
electromagnetic modes (eddy currents in this case), leading to a change in 
energy per area with respect to a uniform bulk medium.

We identify in the two following Sections the mode densities 
$\rho_{\rm CL}( \omega )$ and $\rD( \omega )$
that determine, respectively, the 
free energy due to all modes and due to diffusive modes.
The former quantity is calculated within the Lifshitz
theory for the Casimir effect\cite{lifshitz55}.

\subsection{All modes: Lifshitz mode density}

Let us recall that the mode density $\rho_{\rm CL}( \omega )$
counts how the mode number at a given frequency $\omega$ for 
two half-spaces at separation $L$ differs from the situation of two plates
at infinite distance.
The Lifshitz formula for the Casimir free energy~\cite{Parsegian} can be 
written in the form of Eq.(\ref{eq:free-energy-integral}) so that the
following form of the mode density can be read off
\begin{equation}
	\rCL( \omega ) 
	= -  
	{\rm Im}\,\partial_{\omega}\DispersionD( z=\omega + {\rm i} 0 ).
\label{eq:def-rho-of-omega}
\end{equation}
Here and henceforth, let $z$ denote a complex frequency.
The ``dispersion function'' $\DispersionD( z )$ is given by the integral
\begin{align}
	\DispersionD(z)
	\equiv& \sum_{\sigma=\rmp,\rms}\DispersionD^\sigma \notag \\
	=& \sum_{\sigma=\rmp,\rms}\int\limits_0^\infty
	\!\frac{ k {\rm d} k}{ 2\pi^{2} }
	\log\left[ 1 - r_\sigma^2( \kappa, z ) \, {\rm e}^{- 2 \kappa L} \right],
	\label{eq:def-Lifshitz-dispersion-function}
\end{align}
Here, $\kappa  = \sqrt{ {k}^2 - z^2/c^2 }$, $\sigma$ is a polarization index, 
and $L$ the cavity width. 
In the following, we only consider the s- (or TE-) polarization and drop the
polarization label. The reflection coefficient becomes (using the Drude dielectric 
function~(\ref{eq:drude}))
\begin{subequations}
	\label{eq:def-Lifshitz-rs}
\begin{eqnarray}
r( \kappa, z ) &=& r_{\rm s}( \kappa, z ) 
  = \frac{ \kappa - \sqrt{\kappa^2 + \kappa_{\gamma}^2( z )} }{ \kappa + \sqrt{\kappa^2 +  \kappa_{\gamma}^2( z )} }
  ;
\\
\kappa_{\gamma}( z ) &=& \frac{ \Omega }{ c }
	  \sqrt{  \frac{ z } { z + {\rm i} \gamma } }.
\label{eq:def-kappa-gamma}
\end{eqnarray}
\end{subequations}
All square roots are chosen here with positive real part;
this implies in particular that 
${\rm Im} \,\kappa \le 0$ and
${\rm Im} \sqrt{\kappa^2 + \kappa_{\gamma}^2( z )} \le 0$ for $z$ in the
upper half-plane.

\begin{figure*}[htp] 
   \centering
    \includegraphics[width  = \textwidth]{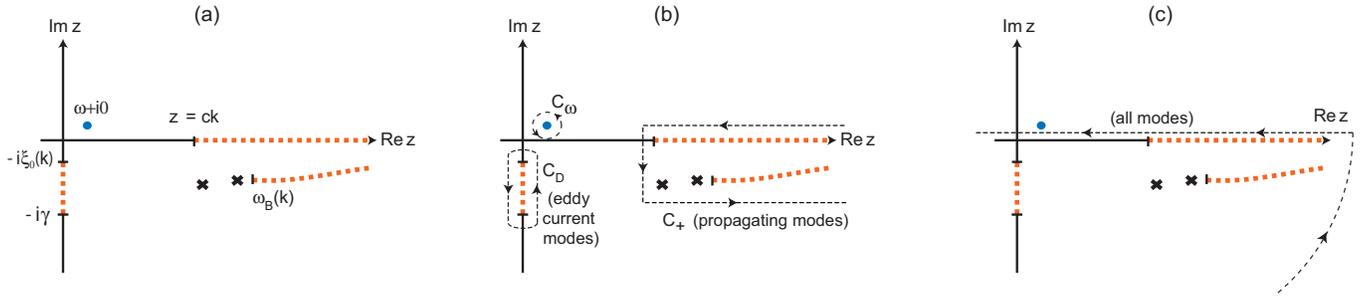}
   \caption{(Color online) (a) Complex eigenfrequencies in the parallel plate geometry, 
   for a fixed wavevector $k$ (not to scale). The 
   thick
   dotted lines show
   branch cuts (three-dimensional mode continuum), crosses mark poles 
   (discrete frequencies). The 
   circular
   dot is a pole 
   of the first factor in Eq.(\ref{eq:contour-cd}). The pole structure
   is symmetric with respect to the imaginary axis, according to
   Eq.(\ref{eq:symmetry-relation}).
   The frequency $\omega_\mathrm{B}(k)$ marks the transition from discrete cavity modes 
   to a continuum of bulk modes (see \cite{intravaia07} for details).
   (b) Integration path: $C_D$ around the eddy current continuum,
   $C_\omega$ around the pole at $z = \omega + {\rm i} 0$,
   $C_+$ around cavity and propagating modes. The corresponding
   paths in the left half-plane are denoted $C_{-\omega}$ and $C_-$,
   respectively.
   (c)
   Integration path that encircles all modes, as relevant for the Lifshitz theory.
   Closing this contour in the upper half-plane, one gets the residue
   from the pole $z = \omega + {\rm i} 0$ in the upper half-plane.}
   \label{ChangePath}
\end{figure*}

\subsection{Eddy current modes}

The dispersion function $\DispersionD( z )$ is analytic in the upper half-plane. 
When it is analytically continued, singularities appear on the real axis
and in the lower half-plane: branch points
where the argument of the logarithm in 
Eq.(\ref{eq:def-Lifshitz-dispersion-function}) vanishes, and branch 
cuts from the square roots involved in the reflection coefficients
(see Fig.\ref{ChangePath}). These
singularities are related to the electromagnetic resonance frequencies of the 
two-plate setup that determine the Lifshitz free energy from the argument
principle~\cite{Schram73, intravaia08, intravaia09b}. They also provide a physically 
well-motivated way to isolate the contribution of a particular class of
modes to the Casimir interaction. 

The eddy current (diffusive) modes, for fixed $k$,
are identified as a branch cut of $\DispersionD(z)$ along the negative 
imaginary 
frequency axis (see figure \ref{ChangePath}), 
$z = -i\xi_0 \ldots - {\rm i} \gamma$ ($\xi_0 = \xi_0( k )$ is defined below).
This branch cut is an example of a dispersion function that is not real
on the imaginary frequency axis, in distinction to the familiar behavior
in the upper half-plane. 
Indeed, one can confirm
from the macroscopic Maxwell equations that 
purely imaginary eigenfrequencies appear in a planar cavity of two 
half-spaces described by the Drude dielectric function \cite{Jackson}.
As is well-known in scattering theory (see, e.g. 
Ref.\cite{Schram73, nesterenko06}),
the branch cut can be interpreted as a dense coalescence of discrete modes,
and the relevant quantity is a mode density given by
\begin{equation}
  \mu_D(\xi) = - {\rm Im}\, \partial_\xi \DispersionD(z=-{\rm i} \xi + 0),
  \quad 0 \le \xi \le \gamma .
\label{eq:def-M-of-xi1}
\end{equation}
The dispersion function is evaluated here to the right of the branch cut.
Continuing analytically from the upper half-plane,
we find that $\kappa$ is mainly real, while
$\kappa_{\gamma}( - {\rm i} \xi + 0) 
= - {\rm i} k_{\gamma}( \xi )$ becomes mainly imaginary with
\begin{equation}
	  k_{\gamma}( \xi ) = \frac{ \Omega }{ c }
	  \sqrt{  \frac{ \xi } {\gamma - \xi } }.
	\label{eq:k-gamma}
\end{equation}
As a consequence, 
$[\kappa^2 - k_{\gamma}^2( \xi )]^{1/2}$ moves to the (negative) 
imaginary axis if $\kappa$ is small enough; more precisely, we require
\begin{equation}
	0 \le \kappa \le k_{\gamma}( \xi )
	\label{eq:identify-branch-cuts}
\end{equation}
This is equivalent to
\be
  \xi_0(k) \leq \xi \leq \gamma
	\label{eq:xi-range}
\ee
where the lower bound $\xi_0(k)$ solves 
\be
  (\gamma-\xi_0)(c^2 k^2+\xi_0^2) = \Omega^2 \xi_0.
	\label{eq:rs}
\ee
We note the limiting behavior $\xi_0( k ) \approx D k^2$ as $k \to 0$
where $D$ is the diffusion coefficient of Eq.(\ref{eq:D}).
In the range (\ref{eq:xi-range}), the reflection coefficient becomes
the unitary number 
\be
  r( \kappa, - {\rm i} \xi + 0) 
  = \frac{ \kappa + {\rm i} \sqrt{  k_{\gamma}^2( \xi ) - \kappa^2} 
  	}{ \kappa - {\rm i} \sqrt{ k_{\gamma}^2( \xi ) - \kappa^2 } }
	\label{eq:r-on-eddy-cut}
\ee
where the sign of the square root applies on the right side of the branch cut
and follows by carefully evaluating 
the imaginary parts of $\kappa$ and $k_\gamma( \xi )$. For imaginary
frequencies outside 
the 
range~(\ref{eq:xi-range}), the reflection coefficient is
real ($-1 < r_{\rm s} < 0$), and the eddy current mode 
density~(\ref{eq:def-M-of-xi1}) vanishes. 

After integrating over $k$, one gets a mode density 
$\mu_D( \xi )$ that is nonzero in the range $0 \le \xi \le \gamma$. 
Finally, the density for eddy current modes $\rD( \omega )$ at
real frequencies is defined by associating to each overdamped mode 
$z = - {\rm i} \xi$ a Lorentzian spectrum centered at zero frequency
whose width is $\sim {\rm Im}\, z$. Referring to
Ref.\cite{intravaia09} for details, we get
\begin{align}
	  \rho_D(\omega)   &= 
	\int\limits_{0}^\gamma 
	  \frac{ \rmd\xi  }{ \pi } \frac{ \xi }{ \xi^2 + \omega^2 } 
	  \mu_D(\xi) .
\label{eq:def-mode-density}
\end{align}

\subsection{Contour integral representation}
\label{s:pole-and-poles}

In this section we derive a contour integral represention for the mode densities 
of the full Casimir-Lifshitz interaction and of the eddy current contribution.
This demonstrates a simple relation between 
$\rCL( \omega )$ and $\rD( \omega )$. We thus prepare the low-frequency
analysis we perform in Sec.\ref{sec:leadingorder}
focusing on the particular case of a good Drude
conductor (i.e., $\gamma \ll \Omega$).

It is easy to see from the sign flip of the root involving 
$\kappa_\gamma( \xi )$ in Eq.(\ref{eq:def-Lifshitz-rs}), 
that the dispersion function $\DispersionD( z )$ jumps and
changes into its complex conjugate
across the branch cut $z = 0 \ldots - {\rm i}\gamma$.
This jump defines the eddy current DOS
in Eq.(\ref{eq:def-M-of-xi1}). The latter can thus be written as 
a contour integral in the complex plane,
\begin{equation}
	  \rho_D(\omega)   = -\,\oint\limits_{C_{D}} 
	\frac{ \rmd z }{2\pi} \frac{z}{z^2-\omega^2}\partial_z \DispersionD(z),	  
	\label{eq:contour-cd}
\end{equation}
where the path $C_{D}$ encircles the cut on the negative imaginary axis
in the positive sense as shown in Fig.\ref{ChangePath}(b).
Now, shifting the contour towards infinity, we encounter the poles at
$z = \pm \omega$ from Eq.(\ref{eq:contour-cd}) and other singularities
(poles and branch cuts) of $\partial_z \DispersionD(z)$. 
The behavior of the exponential ${\rm e}^{ - 2 \kappa L }$ for 
$|z| \to \infty$ makes $\DispersionD(z)$ vanish at infinity. 
Hence we conclude that the integral around $C_D$ is equal to 
the negative residues of the poles $z = \pm \omega$ 
minus integrals over the contours $C_\pm$ in Fig.\ref{ChangePath}(b)
that encircle the singularities near the left and right half of the real 
axis.
We use here the link between the dispersion function and the 
response (or Green) function of the two-plate cavity 
\cite{Davies72, Schram73} 
that entails the symmetry relation
\be\label{eq:symmetry-relation}
  \DispersionD(-z^{*}) = \DispersionD^{*}(z)
\; .
\ee
As a consequence, complex mode frequencies and branch cuts appear in 
pairs on opposite sides of the imaginary axis. (In Fig.\ref{ChangePath}, 
only the right half is shown.)

The residues at the poles are easily calculated from the 
contours $C_{\pm \omega}$ in Fig.\ref{ChangePath}(b):
\begin{align}
\oint\limits_{C_{\omega}+C_{-\omega}} 
  \frac{ {\rm d}z }{2\pi} &
  \frac{z}{z^2-\omega^2}\partial_z \DispersionD(z)
	  =\frac{{\rm i}}{2}\left(\partial_{z}\DispersionD(\omega)+
	  \partial_{z}\DispersionD(-\omega)\right)\notag\\
	  &=-{\rm Im}\partial_{\omega}\DispersionD(\omega)=\rCL(\omega)
	  \label{eq:final-tildeM(0)/0}
\end{align}
We thus recover the mode density for the Lifshitz theory
as one term in the eddy current DOS. This is actually not surprising,
since $\rCL( \omega )$ can be written as a similar contour integral
as Eq.(\ref{eq:contour-cd}), but evaluated along a contour just above
the real axis [Fig.\ref{ChangePath}(c)]
and closed at infinity in the lower half plane.
This contour encircles all singularities of the dispersion function 
as it should, since the Lifshitz theory accounts for all modes.
If this contour is shifted through infinity into the upper half-plane,
only the two residues
calculated in Eq.(\ref{eq:final-tildeM(0)/0}) contribute since the
dispersion function is analytic inside the contour.

In conclusion, we can write the following splitting of the mode density
for the Casimir effect
\begin{equation}
	\rCL( \omega ) = \rD( \omega ) + \rho_{\pm}( \omega )
	\label{eq:eddies-vs-CasimirLifshitz}
\end{equation}
where the last term gives the contribution of modes near
the real axis (contour $C_{+}$ in Fig.\ref{ChangePath}(b), and corresponding $C_-$ in the left half-plane).
By continuity with the limiting case $\gamma \to 0$,
we can identify the latter modes with propagating modes in the vacuum cavity,
in the bulk, or with electromagnetic surface modes (for example, surface
plasmons that appear in the TM-polarization).
We shall see in the next section
that for nonzero, but small $\gamma \ll \Omega$,
the mode density 
$\rho_{\pm}( \omega )$ becomes small 
at low frequencies ($\omega < D/L^2, \gamma, c/L$),
so that in this range, 
the full electromagnetic DOS $\rCL( \omega )$ nearly coincides with the
eddy current DOS $\rho_{D}( \omega )$.

\section{Low-frequency expansion}
\label{sec:leadingorder}

We calculate now the small $\omega$ expansion of
the density of states for eddy current modes. According to 
Eq.\ (\ref{eq:eddies-vs-CasimirLifshitz}), we start with the full 
Casimir-Lifshitz interaction and discuss then the differences between
the two. We begin with a general estimate of the scaling for good
conductors.

\subsection{Scaling for weak damping}
\label{s:estimate-scaling}

The analysis in the complex plane, as illustrated in Fig.\ref{ChangePath},
suggests that the density of diffusive modes is concentrated in a range
$\sim \gamma$ near zero frequency. Anticipating from the analysis below
a total number of $\sim 1/L^2$ modes per unit area, one gets for $\gamma\to 0$ a scaling
$\rho_D( \omega ) \approx 1/(\gamma L^2) \theta( \omega / \gamma, 
L/\lambda )$
where the frequency appears in the 
function $\theta$ only in the dimensionless form $\omega / \gamma$
(and similarly for the distance $L / \lambda \equiv \Omega L / c$).
A different behavior emerges for propagating modes (inside the contour 
$C_+$ in Fig.\ref{ChangePath}): they move onto the real axis for small
$\gamma$ and contribute to $\rho_{\pm}( \omega )$ in the range
$\omega \sim c/L, \Omega$. Their contribution at much lower frequencies
that interests us here, is proportional to their imaginary part and therefore
scales linearly with $\gamma$. This observation shall provide us with a 
simple rule to identify the respective contributions of eddy current and
propagating modes in the full (Lifshitz) mode density, Sec.\ref{s:Lifshitz-rho-0}.
Note that we consider here the case of a fixed (temperature-independent)
scattering rate $\gamma$.

Some further corroboration of these rough estimates is desirable.
Let us consider for the simplicity of argument that the dispersion function 
$\partial_z \DispersionD( z )$ has only discrete poles $\omega_n( k )$
in the lower half-plane, labelled by the momentum quantum number $k$.
This can be achieved by enclosing the system in a finite 
box \cite{Schram73, intravaia08}. One
recovers the branch cuts by taking the box size to 
infinity~\cite{nesterenko06}.
From the symmetry relation~(\ref{eq:symmetry-relation}), the poles 
occur either on the imaginary axis (as for diffusive modes) or pairwise in the
lower left and right quadrants (as for propagating modes). The two
terms $\rD( \omega )$ and $\rho_{\pm}( \omega )$ collect 
these poles, respectively. 

We make the replacement ${\rm d}\xi\, \mu_D( \xi ) \mapsto
{\rm d}^2k/(2\pi)^2 \sum_{n\in \text{eddy}}$
and find that 
the DOS for diffusive modes $\rho_D( \omega )$ can be written in the 
following scaling form
\begin{equation}
	\rD( \omega ) = \frac{ 1 }{ \gamma }
	\int\!\frac{ k {\rm d}k }{ 2 \pi^2 }
\,\bigg[\sum_{n\in \text{eddy}}
\frac{ \xi_n(k)/\gamma }{ 
(\omega / \gamma) ^{2} + (\xi_n(k)/\gamma)^{2}
}
\bigg]
^{L}_{L\to \infty}\nonumber\\
	\label{eq:re-scale-diffusive-DOS}
\end{equation}
where the limit of the mode branches for two separate plates 
($L \to \infty$) is subtracted. 
We have already seen that the mode frequencies satisfy
$\xi_n( k ) \le \gamma$. As a consequence, 
the integral tends toward a finite limit
as $\gamma \to 0$,
$\rD( \omega )$ depends only on the scaled
frequency $\omega / \gamma$ and is proportional to 
the scaling factor $1/\gamma$. This implies in particular that the
integral over the diffusive mode density 
can give a nonzero contribution even as $\gamma \to 0$. We confirm
these results in Eq.(\ref{eq:coefficientResult}) below.

The density of propagating modes $\rho_{\pm}( \omega )$ shows a
different scaling with $\gamma$. With the same re-writing, the contours
$C_{+}$ and $C_{-}$ collect the modes away from the imaginary axis
and lead to the representation
\begin{align}
	\rho_{\pm}( \omega ) &= 
	-  \oint_{C_+ + C_-}
	\!\frac{ {\rm d}z }{ 2\pi } \frac{ z }{ z^2 - \omega^2 }
	\partial_z \DispersionD( z )
	\notag
	\\
	& = 
	-
	\int_0^\infty\!\frac{ k {\rm d}k }{ 2\pi^2 }
\,{\rm Im}\bigg[\sum_{n\in \text{prop}}\frac{\omega_n(k)}{\omega^{2}-\omega^{2}_n(k)}
\bigg]^{L}_{L\to \infty}
	\label{eq:propagating-DOS-as-sum}
\end{align}
where in the second line,
we represent the modes by the poles in the lower right quadrant.
Now, the imaginary part of the eigenfrequency $\omega_n( k )$
is negative and, for a small Drude scattering rate,
of the order $\gamma$. A typical scale for its real part, on the other
hand, is the lowest cavity eigenfrequency $c/L$ or the plasma frequency 
$\Omega$. Although we do not need them here, recall that
the surface plasmon modes appear at $\approx \Omega / \sqrt{ 2 }
- {\rm i} \gamma / 2$ for $k \gg \Omega / c$~\cite{Raether}.
For an estimate of $\rho_{\pm}( \omega )$, we focus on frequencies
$\omega$ much smaller than the real part, $\omega \ll c/L, \Omega$ 
and take $1/L$ to estimate the relevant wave vectors. This gives
\begin{equation}
	\omega \to 0: \quad
	\rho_{\pm}( \omega ) \sim 
	\mathcal{O}\big( \frac{ \gamma }{ \Omega^2 L^2 } \big) 
	\ldots 
	\mathcal{O}\big( \frac{ \gamma }{ c^2 } \big)
	\label{eq:estimate-propagating-DOS}
\end{equation}
as we confirm in Eq.(\ref{eq:calc-Mgamma-diffusive}) below.
This small ``tail'' of the mode density near zero frequency can be understood
from the broadening of the discrete modes due to damping (a $\delta$-peak
becomes similar to a Lorentzian, see also Ref.\cite{Hanke95}). In particular,
it vanishes in the limit $\gamma \to 0$ where $\rho_{\pm}( \omega )$
goes over into the mode density of the plasma model which scales proportional
to $\omega^{2}$.

To summarize this estimate, we expect from 
Eq.(\ref{eq:eddies-vs-CasimirLifshitz}) that as $\gamma \to 0$,
the low-frequency mode density for the 
Casimir-Foucault interaction and for the full Casimir interaction coincide
in order $1/\gamma$, 
with a small difference $\sim \gamma$
arising from propagating modes. 
These contributions are calculated in the following sections. 
We are thus able to check 
our approach against the free energy expansion of 
Refs.\cite{hoye07,ellingsen08}.

\subsection{Lowest order: Lifshitz theory}
\label{s:Lifshitz-rho-0}

Let us calculate 
the coefficients $\coeffM^{(1)}_{D}$ and $\coeffM^{(3/2)}_{D}$ defined in Eq.\ \eqref{eq:small-omega-expansion}, starting with the leading order in the small parameter $\gamma/\Omega$ which, as we have just seen, is provided by the 
full Lifshitz theory. It is convenient to work with the integrated mode density 
which from Eq.(\ref{eq:def-rho-of-omega}), we can read off as
\begin{equation}
	\intMD_{\rm CL}( \omega ) 
	= {\rm Im}\,\DispersionD( \omega + {\rm i}0 ).
	\label{eq:Lifshitz-M-of-omega}
\end{equation}
We therefore start by analyzing in detail the behavior of $\DispersionD(z)$ 
for small frequencies $|z| \ll \gamma$. 

The $k$-integral in Eq.(\ref{eq:def-Lifshitz-dispersion-function}) is re-written
in terms of a real variable $y > 0$ defined by
$\kappa = y \kappa_{\gamma}(z)$. This is equivalent to
a shift of the integration path in the complex $\kappa$-plane
along a 
more convenient direction: one still has convergence 
from the exponential
$\exp( - 2 \kappa L )$ because ${\rm Re}\, \kappa_\gamma( z ) > 0$
(keeping clear of the branch cut for $z$ on the negative imaginary axis).
The reflection coefficient~(\ref{eq:def-Lifshitz-rs}) becomes real along this
direction and independent of $z$:
\begin{equation}
r( y ) =
  \frac{ y - \sqrt{ y^2 + 1 } 
  	}{ y + \sqrt{ y^2 + 1 } } = 	\frac{ 1 }{ (y + \sqrt{ y^2 + 1 })^2 } .	
	\label{eq:Simens-trick-for-rs}
\end{equation}
We get
\begin{equation}
  \DispersionD( z )
  =
  \kappa^{2}_{\gamma}(z)
  \int\limits_{\chi( z )}^{\infty} \frac{y \rmd y}{2\pi^{2}} 
  \log\left[1 - r^2(y) e^{-2y\kappa_{\gamma}(z) L} \right]
	\label{eq:MCL-integrals}
\end{equation}
where the lower limit is given by the complex number
\begin{equation}
	  \chi(z) =   -\frac{\rmi z}{c \kappa_{\gamma}(z)} = 
	  - \frac{ {\rm i} z }{ \Omega } \sqrt{ \frac{ z + {\rm i} \gamma }{ z } }.
	\label{eq:lower-limit-CL}
\end{equation}
For $|z| < \gamma$, we have $|\chi( z )| \le \sqrt{2} (\gamma/\Omega)
\ll 1$ for a good conductor, and to leading order, we can replace the
lower limit in Eq.(\ref{eq:MCL-integrals}) by zero. This defines 
$\DispersionD_0( z )$ and via 
Eq.(\ref{eq:Lifshitz-M-of-omega}), 
$\intMD_{{\rm CL},0}( \omega )$. We write 
$\intMD_{{\rm CL},\gamma}( \omega )$ for
the error (i.e., the integral from $0$ to $\chi( z )$) and calculate it in
Eq.(\ref{eq:Mg-cubic-T}) below.

By inspection, $\DispersionD_0( z )$ depends on $z$ only via the function
$\kappa_\gamma( z )$ that can be written as 
\begin{equation}
  \kappa_\gamma( z ) = 
  \kappa_{1}(z / \gamma ) = 
  \frac{\Omega}{c}\sqrt{\frac{ z / \gamma }{z / \gamma + \rmi}}
  ,
	\label{eq:def-kappa-1}
\end{equation}
involving the scaled quantity $z / \gamma$. 
From the reflection coefficient $r( y )$, the relevant integration domain
is $0 \le y\lesssim 1$. We can therefore expand the exponential 
in Eq.(\ref{eq:MCL-integrals}) 
provided $\kappa_1( z / \gamma ) L \ll 1$. This yields
the condition $(|z| / \thouless)^{1/2} \ll 1$ where $\thouless$ is the 
Thouless frequency
introduced in Eq.(\ref{eq:D}). Expanding to the first order in
this small parameter, we get ($D$ is the diffusion coefficient)
\begin{align}
	\DispersionD_0( z ) &\approx
	\frac{ z }{ {\rm i} D }
	\int\limits_{0}^{\infty} 
	\frac{y \rmd y}{2\pi^{2}}\log\left[ 1 - r^2(y) \right]
	\label{eq:final-M(0)}\\
	  &
	  {} +\frac{ z }{ {\rm i} D }
	  \Big( \frac{ z }{ {\rm i} \thouless } \Big)^{1/2}
	  \int\limits_{0}^{\infty} 
	\!
	\frac{y^{2} \rmd y}{2\pi^{2}} \frac{2 r^2(y) }{1 - r^2(y)} 
	+ \mathcal{O}^{2}( z / \thouless )
	\nonumber
\end{align}
where powers $z$ and $z^{3/2}$ have appeared.
The integrals can be solved exactly (see Appendix \ref{a:Lifshitz-integrals}),
and we get from Eq.(\ref{eq:Lifshitz-M-of-omega})
the following approximation to the Lifshitz integrated DOS
\begin{equation}
\intMD_{{\rm CL},0}(\omega)
\approx
\frac{2\log 2 - 1}{8\pi^{2}} \frac{\omega}{D} - \frac{L\sqrt{2}}{24\pi^{2}}
\left(\frac{\omega}{D}\right)^{3/2}
	\label{eq:coefficientResult}
\end{equation}
valid for $\omega \ll \thouless, \gamma$. This proves the first term
in Eqs.(\ref{eq:result-rhoD}).
It is clear from this calculation (a power series in $(\omega / \thouless)^{1/2}$)
that these results cannot be applied for $\gamma\to 0$ at fixed 
$\omega > 0$. In other words, the limits $\gamma \to 0$ and 
$\omega\to 0$ do not commute. 
For a discussion, see Refs. \cite{intravaia08, ellingsen08b}.

Calculate now the small correction 
$\intMD_{{\rm CL},\gamma}( \omega )$ from the lower integration limit
in Eq.(\ref{eq:MCL-integrals}). This arises between the boundaries
$y = 0$ and $y = \chi( z )$. Recall that in the limit $\gamma \ll \Omega$,
we have $\vert \chi(z)\vert\ll 1$ and expand the integrand for 
$y\ll 1$. This gives 
\begin{align}
\intMD_{{\rm CL},\gamma}( \omega )
& \approx 
	-
	{\rm Im}\,
	\kappa^{2}_{\gamma}(z)\!\!
	\int\limits_{0}^{\chi(z)} \frac{y \rmd y}{2\pi^{2}} 
	\log\left[2y(2+\kappa_{\gamma}(z) L) \right]\notag\\
	& \approx
	-\frac{ 1 }{16\pi}\frac{ \omega^{2} }{ c^{2} } 
	,
	\label{eq:Mg-cubic-T}
\end{align}
one half the mode density for the so-called plasma model
where $\gamma = 0$ is taken from the outset.
Notably, this term gives a contribution to the free energy proportional to 
$T^{3}$, which exactly coincides with the expression given in \cite{ellingsen08,ellingsen09}. Note that the term scaling with 
Eq.(\ref{eq:estimate-propagating-DOS}) has not appeared in the full
(Lifshitz) mode density. We outline an interpretation in Sec.\ref{s:discussion}.

\subsection{Eddy current modes}

We now address the density of eddy current modes alone
that involves according to Eq.(\ref{eq:def-mode-density})
an integral along the branch cut of $\DispersionD( z )$
on the imaginary axis. It is again convenient to work out the
integrated mode density ${\cal M}_D( \omega )$.
A partial integration leads to the integral representation
\begin{equation}
\intMD_{D} (\omega)= 
-\int\limits_{0}^\gamma  \frac{ \rmd\xi }{ \pi } 
\frac{ \omega }{ \xi^2+\omega^{2} } M_D(\xi).
	\label{eq:rho-of-zero-0}
\end{equation}
where $M_D(\xi)$ is the integrated mode density along the branch cut.
By changing the momentum variable from $k$ to $\kappa$, this function 
can be written as 
\begin{equation}
	  M_D(\xi) = -
	  \int\limits_{\xi/c}^{k_{\gamma}(\xi)} 
	  \frac{\kappa \rmd \kappa}{2\pi^2} 
	  {\rm Im} \log[1 - r^2( \kappa, - {\rm i} \xi ) e^{-2\kappa L}],
	\label{eq:def-M-of-xi}
\end{equation}
where the integrand is zero above the upper integration limit
$k_\gamma(\xi)$ that was defined in (\ref{eq:k-gamma}).

The limiting behavior of this expression for a good conductor can be
worked out similar to Eq.(\ref{eq:MCL-integrals}).
Writing the integral in terms of $x = \xi / \gamma$, we see that 
${\cal M}_D( \omega )$ [Eq.(\ref{eq:rho-of-zero-0})] 
depends on the scaled frequency $\omega / 
\gamma$. The upper integration limit takes a form similar 
to Eq.(\ref{eq:def-kappa-1}),
\begin{equation}
	k_\gamma( \xi ) = k_{1}( x )
	= \frac{\Omega}{c}\sqrt{\frac{x}{1-x}}
	\label{eq:def-k-1}
	, 
\end{equation}
while the lower one, $\xi / c = x \gamma / c$, can be taken as 
small compared to
the typical values $\kappa \sim 1/L$ and $\kappa \sim k_{1}( x )$
that appear in the integrand.

This motivates again a splitting of $M_D( \xi )$ in two terms,
a first one where the lower boundary in Eq.(\ref{eq:def-M-of-xi})
is taken as zero, and a correction, similar to what we did 
after Eq.(\ref{eq:MCL-integrals}).
The two terms produce a split of the mode density~(\ref{eq:rho-of-zero-0})
into 
\begin{subequations}
\begin{equation}
	\intMD_{D,0}( \omega ) + \intMD_{D,\gamma}( \omega ).
	\label{eq:decomposition}
\end{equation}
where the first term can be written as
\begin{align}
\intMD_{D,0}( \omega )
         &=\int\limits_{0}^1
         \frac{ {\rm d}x }{ \pi }
	  \frac{ \omega/\gamma }{x^{2} + (\omega / \gamma)^{2}} 
	  \label{eq:decomposition-R0}
\\
	& {} \times
	  \int\limits_{0}^{k_{1}(x)} 
	  \frac{\kappa\, \rmd \kappa}{2\pi^2} 
	  {\rm Im}
	  \log[1 - r^2( \kappa, - {\rm i}x \gamma ) e^{-2\kappa L}]
	  \notag
\end{align}
\end{subequations}
Here, we have succeeded in removing from the integrand all dependence
on $\gamma$, except for the frequency scaling. The second term,
$\intMD_{D,\gamma}( \omega )$, is discussed in 
Sec.\ref{s:cavity-correction}, Eq.(\ref{eq:calc-Mgamma-diffusive}).
This term is related to the correction proportional to $\gamma$ identified
in Sec.\ref{s:estimate-scaling}, the only difference being that we are dealing
here with integrated mode densities. 
The expression $\intMD_{D,0}( \omega )$ [Eq.(\ref{eq:decomposition-R0})]
is nonzero in the limit $\gamma \to 0$, except for the appearance of the
scaled frequency $\omega / \gamma$. Therefore, this term
corresponds to the (differential) mode density scaling with $1/\gamma$ 
of Sec.\ref{s:estimate-scaling}. 
Since we know from Eq.(\ref{eq:eddies-vs-CasimirLifshitz})
that the leading orders for $\gamma \to 0$ coincide for the diffusive
modes and the Lifshitz theory, we can conclude
\begin{equation}
\intMD_{D,0}( \omega ) = 
\intMD_{{\rm CL},0}( \omega )
	\label{eq:lowest-order-eddies=Lifshitz}
\end{equation}
provided the frequency $\omega$ is below the range where other
(propagating) modes appear that are not contained in 
$\intMD_{D}( \omega )$.
The identity~(\ref{eq:lowest-order-eddies=Lifshitz})
is checked by a direct calculation in Appendix~\ref{a:direct-calc}.

\subsection{Damping correction of eddy current modes}
\label{s:cavity-correction}

We now show that one gets for good conductors the second term,
of relative order $(\gamma/\Omega)^2$, in the coefficient
$\coeffM_D^{(1)}$ [Eq.(\ref{eq:M1})]. It arises from
the correction $\intMD_{D,\gamma}( \omega )$
to the diffusive mode density. 
It is interesting that this shows a scaling $\sim \gamma \omega$,
in distinction to the correction in the Lifshitz theory, 
Eq.(\ref{eq:Mg-cubic-T}).

The second term in Eq.(\ref{eq:decomposition}),
$\intMD_{D,\gamma}( \omega )$, is of the same form 
as Eq.(\ref{eq:decomposition-R0}), with the upper limit $k_1( x )$
replaced by $\gamma x / c$.
For good conductors, 
the upper integration limit
$\kappa \le \gamma x / c$ is small compared to the
scale $k_1( x )$ [Eq.(\ref{eq:def-k-1})] that appears in the reflection
coefficient. The argument of the
exponential is small if we take $\gamma \ll c / L$. Expanding both
quantities for small $\kappa$, we get
\begin{align}
\intMD_{D,\gamma}( \omega )
         &\approx \int\limits_{0}^1
         \frac{ {\rm d}x }{ \pi }
	  \frac{ \omega/\gamma }{x^{2} + (\omega / \gamma)^{2}} 
\\
	& {} \times
	  \int\limits_{0}^{\gamma x/c} 
	  \frac{\kappa\, \rmd \kappa}{2\pi^2} 
	  {\rm Im}
	\log[ 
	2 \kappa ( L + 2 {\rm i} / k_1(x) ).
	]
	  \notag
\end{align}
The imaginary part does not depend on $\kappa$, and the integration gives
a factor $\frac12 (\gamma x/c)^2$.
At this stage, we can take the low-frequency limit ($\omega \ll \gamma$) 
and are left with
\begin{align}
\intMD_{D,\gamma}( \omega )
  & \approx 
  \frac{\omega\gamma}{4\pi^2 c^2}
  \int_0^1 \frac{ \rmd x }{ \pi }
  \arctan\left(\frac{2 \lambda}{ L}
  \sqrt{\frac{1-x}{x}}\right)\notag \\
  & = \frac{\omega}{D}\frac{\gamma^2}{4\pi^2 \Omega^2}\frac{\lambda}{2\lambda + L},
	\label{eq:calc-Mgamma-diffusive}	
\end{align}
where $\lambda = c/\Omega$ is the plasma wavelength.
This yields the correction to $\coeffM_D^{(1)}$ appearing in
Eq.\ (\ref{eq:M1}). We have checked that 
$\intMD_{D,\gamma}( \omega )$ does not contain, at the next order,
a fractional power $\omega^{3/2}$, as found for $\intMD_{D,0}( \omega )$.

We suggest the following interpretation for this correction: it is related to
the mutual influence of the two types of modes, overdamped and propagating
waves. To wit, as the two slabs
approach each other, the different mode frequencies cannot shift independently 
because taken all together, they have to satisfy a sum rule quoted in 
Ref.\cite{intravaia08}:
\begin{equation}
	\int\!{\rm d}^2k 
	\,\, 
	\left[\sum_{\text{all 
	modes }}{\rm Im}\,\,
	\omega_n(k)  \right]^{L}_{L\to\infty}
	= 0
	\label{eq:sum-rule}
\end{equation}
where the notation assumes that branch cut continua have been discretized
(see Sec.\ref{s:estimate-scaling}).
The eddy current modes play a crucial role in satisfying this sum rule. Indeed, 
any modification in the imaginary part of the propagating (cavity and bulk) modes 
due to a change of the distance $L$ (i.e. the propagating modes leave
the continuum above the plasma frequency and become discrete cavity modes 
as the distance $L$ is increased)
is simultaneously balanced by a shift in the diffusive mode density on 
the imaginary axis that extends down to $- {\rm i} \gamma$. 

Due to the sum rule~(\ref{eq:sum-rule}), the small correction for eddy
currents appears also, with opposite sign, in the propagating modes. 
For this reason, the Lifshitz mode density does not contain this term
[see Eq.(\ref{eq:coefficientResult})], and its
next-order correction Eq.(\ref{eq:Mg-cubic-T}) is independent of the damping 
rate $\gamma$. 

\section{Discussion and conclusions}
\label{s:discussion}

We have calculated the low-temperature behavior of the interaction between 
two parallel half-spaces 
across a gap of width $L$ due to low-frequency Johnson noise in the bulk
of the conducting medium, in particular eddy or Foucault currents that
are coupled to TE-polarized electromagnetic fields.
The interaction is calculated in orders $T^2$ and $T^{5/2}$ and is compared
to the Casimir free energy within the Lifshitz theory for Drude metals.
A striking result is uncovered: the low-temperature correction to the 
Casimir effect between parallel slabs of good Drude conductors 
is dictated entirely by the contribution from eddy currents, as demonstrated
by the two leading order correction terms as $T\to 0$. This adds a further
piece of support to the findings of Ref.\cite{intravaia09} where the unusual
physics of the thermal Casimir effect between Drude conductors is attributed
to the interaction between eddy currents. 

Within our approach, we find small differences to the free energy
that are of second order in the ratio 
scattering rate to plasma frequency, $\gamma / \Omega $ 
[Eq.(\ref{eq:M1})]. This correction reflects the
mutual influence between the modes that are constrained by a sum
rule, Eq.(\ref{eq:sum-rule}). Note the curious fact that this makes
$\Delta F_D$ depend on $L$ already at order $T^2$, different from
$\Delta \FCL$.
Therefore the eddy current 
interaction makes a tiny contribution to the pressure ($p= -\partial F / \partial L$) 
quadratic in temperature. However this is exactly cancelled by a corresponding 
contribution from propagating modes and the resulting Casimir pressure is 
proportional to $T^{5/2}$ to leading order, as Eq.\ \eqref{eq:coefficientResult} shows.

Let us finally emphasize the analysis of the singularities
of the Lifshitz dispersion function that we performed in the complex plane.
This picture identifies in a natural way the mode frequencies of the system,
even in the presence of dissipation, and 
justifies a natural splitting of the free energy in contributions of specific types
of modes.
We gained in particular the insight that the mode density 
for the full Casimir-Lifshitz interaction is simply the sum of
eddy current modes and of propagating cavity and bulk modes. The second
contribution becomes small at low frequencies, 
weak damping and not too large distances
because the complex mode frequencies are located sufficiently far away from 
the origin. This provides a deeper understanding why propagating modes are of 
little relevance to the temperature dependence of the Casimir-Lifshitz interaction 
between Drude metals. Indeed, this dependence was previously found to originate
primarily in low-frequency evanescent modes \cite{torgerson04, svetovoy07}.

\subsection*{Acknowledgement}

We have benefited from discussions with Gert-Ludwig Ingold. Support from the European Science Foundation (ESF) within the activity
`New Trends and Applications of the Casimir Effect'
(\texttt{www.casimir-network.com}) is greatfully acknowledged. FI acknowledges partial financial support by the Humboldt foundation and LANL.

\appendix

\section{Integrals for Lifshitz theory}
\label{a:Lifshitz-integrals}

The integrals in Eq.(\ref{eq:final-M(0)}) can be evaluated with the
substitution $y = \sinh t$. This simplifies the reflection 
coefficient~(\ref{eq:Simens-trick-for-rs}) into $r(y) = - {\rm e}^{-2t}$. 
Hence
\begin{equation}
	\int\limits_{0}^{\infty} 
	\rmd y\, y \log\left(1 - r^2(y) \right)
	= 
	\int\limits_{0}^{\infty} 
	\rmd t\, \frac{ \sinh 2t }{ 2 }
	\log\left(1 - {\rm e}^{-4t}\right)
	\label{eq:substitute-D1}
\end{equation}
Expanding the logarithm, integrating term by term and evaluating
the sum, we get
\begin{align}
  &\int_0^\infty \frac{ \sinh 2t }{ 2 } \log(1 - {\rm e}^{-4t})
=-\frac{ 2\log 2-1 }{4}
.
	\label{A2}
\end{align}
For the second integral in Eq.(\ref{eq:final-M(0)}), the same substitution gives
\begin{equation}
	\int\limits_{0}^{\infty} 
	\!\rmd y\,
	\frac{2 y^{2} r^2(y) }{1 - r^2(y)} 
=
	\int\limits_{0}^{\infty} 
	\rmd t\, 
	\frac{ \sinh t \, {\rm e}^{-2t} }{ 2 } = \frac{ 1 }{ 6 }
	.
	\label{eq:result-order-3/2}
\end{equation}

\section{Integrals for eddy currents}
\label{a:direct-calc}

We prove here Eq.(\ref{eq:lowest-order-eddies=Lifshitz}): the low-frequency
mode densities for eddy currents, ${\cal M}_{D,0}( \omega )$,
and for all modes, ${\cal M}_{{\rm CL},0}( \omega )$,
coincide to leading order in $\gamma$.

Consider Eq.(\ref{eq:decomposition-R0})
for the eddy current mode density. We want to write this as a contour
integral, similar to Eq.(\ref{eq:contour-cd}),
around the eddy current branch cut $C_D$ in Fig.\ref{ChangePath}.
Note first that the ${\rm Im}$ can be pulled in front of 
the $\kappa$-integral and that integral be extended from
$k_1( x )$ to $\infty$. This is possible without changing the value of
the integral if $\kappa$ is taken just below the real axis, 
the reflection coefficient~(\ref{eq:r-on-eddy-cut}) getting real and smaller 
than unity in modulus. Hence, the logarithm is real, and this part of the
integration range does not make any contribution to the imaginary part.

The contour integral in the variable $z = - {\rm i} \gamma x \pm 0$
finally takes a form similar to Section~\ref{s:pole-and-poles}
\begin{equation}
	{\cal M}_{D,0}( \omega ) = 
	- \oint_{C_D} \!\frac{ {\rm d}z }{Ê2\pi }
	\frac{ \omega }{ \omega^2 - z^2 }
	{\cal D}_{D,0}( z )
	\label{eq:eddy-DOS-as-contour-again}
\end{equation}
where ${\cal D}_{D,0}( z )$ is the integral
\begin{equation}
	{\cal D}_{D,0}( z ) = 
	\int\limits_0^\infty\!\frac{ \kappa \, {\rm d}\kappa }{ 2 \pi^2 }
	\log[ 1 - r^2( \kappa, z) {\rm e}^{ - 2 \kappa L }] 
	,
	\label{eq:def-eddy-dispersion-function}
\end{equation}
and the reflection coefficient is given by Eq.(\ref{eq:def-Lifshitz-rs}).
Note that to the right of the branch cut,
$k_1( x ) = {\rm i} \kappa_\gamma( z )$.

The variable change $\kappa = y \kappa_\gamma( z )$
with $y \ge 0$ now shows that the function ${\cal D}_{D,0}( z )$ 
is indeed identical to the small-$\gamma$
approximation to the Lifshitz dispersion function,
${\cal D}_{0}( z )$, defined by setting
the lower integration limit in Eq.(\ref{eq:MCL-integrals}) to zero. 
Note that this actually shifts the integration path in the lower right
quadrant of the complex $\kappa$-plane: from ${\rm Re}\,
\kappa_\gamma( z ) > 0$, convergence at $y \to \infty$ is
secured. The reflection coefficient $r(y)$~[Eq.(\ref{eq:Simens-trick-for-rs})]
is analytic and of modulus smaller than unity in this quadrant, hence the
logarithm encounters no branch points.

We still have to evaluate the integral~(\ref{eq:eddy-DOS-as-contour-again})
and do this with the same technique as in Sec.\ref{s:pole-and-poles}. Pulling 
the contour $C_D$ to infinity, one encouters simple poles at 
$z = \pm \omega$. In the present
case, we can argue that
the function ${\cal D}_{0}( z )$ is analytic in the right half-plane
and by the symmetry relation~(\ref{eq:symmetry-relation}) also in the
left half-plane: this is due to the way the integration variable $y$ keeps
the wave vector $\kappa$ clear of the branch cuts that are located inside 
the contours $C_{\pm}$ [Fig.\ref{ChangePath}]. Indeed, across these cuts 
either
$\kappa$ or $\sqrt{ \kappa^2 + \kappa_\gamma^2( z ) }$ are purely
imaginary and jump in sign. This never happens along the path parametrized
as $\kappa = y \kappa_\gamma( z )$, as can be checked easily. 
Indeed, if $z$ is in the right
half-plane, $\kappa$ remains in the lower right quadrant, excluding the real
and imaginary axes.

As a consequence of ${\cal D}_{0}( z )$ being analytic in the left and right 
half-planes,
the integral~(\ref{eq:eddy-DOS-as-contour-again}) is given by the
pole contributions ${\cal D}_{0}( \pm\omega )$ only. Referring 
to Eq.(\ref{eq:Lifshitz-M-of-omega}), we thus get 
the desired link to the approximated Lifshitz mode density 
\begin{equation}
	{\cal M}_{D,0}( \omega ) = 
	{\rm Im}\, {\cal D}_{0}( \omega ) = 
	{\cal M}_{{\rm CL},0}( \omega )
	\label{eq:prove-eddy=Lifshitz}
\end{equation}
which is Eq.(\ref{eq:lowest-order-eddies=Lifshitz}).

%

\end{document}